\documentclass[runningheads]{llncs}

\usepackage[T1]{fontenc}
\usepackage{todonotes}
\usepackage{siunitx,booktabs}
\usepackage[misc,geometry]{ifsym}
\usepackage{hyperref}
\usepackage{amssymb}
\usepackage{graphicx}
\usepackage{pifont}
\newcommand{\cmark}{\ding{51}}%
\newcommand{\xmark}{\ding{55}}%
\usepackage{pdflscape}
\usepackage{float}
\usepackage{comment}
\usepackage{svg}
\usepackage{microtype}
\usepackage{array}
\usepackage{ragged2e}
\usepackage{makecell}
\usepackage{hyperref}

\usepackage{tikz}
\usetikzlibrary{shapes.geometric,arrows,positioning}
\usetikzlibrary{calc} 

\tikzstyle{startstop} = [rectangle, rounded corners, minimum width=3cm, minimum height=1cm,text centered, draw=black, fill=blue!20]
\tikzstyle{process} = [rectangle, minimum width=3cm, minimum height=1cm, text centered, draw=black, fill=white!20]
\tikzstyle{arrow} = [thick,->,>=stealth]

\usepackage{color}

\definecolor{customblue}{HTML}{548dc0}
\definecolor{customgreen}{HTML}{4cce7c}
\definecolor{customred}{HTML}{ff6a4c}
\definecolor{customorange}{HTML}{ffb44c}

\begin{document}

\title{TAPAS: A Pattern-Based Approach to Assessing Government Transparency}

%
%
\author{Jos Zuijderwijk\inst{1,2}\textsuperscript{(\Letter)}\orcidID{0009-0008-9561-9354} \and Iris Beerepoot\inst{1}\orcidID{0000-0002-6301-9329} \and Thomas Martens\inst{1}\orcidID{0009-0007-2353-7301} \and Eva Knies\inst{1}\orcidID{0000-0003-0918-7572} \and Tanja van der Lippe\inst{1}\orcidID{0000-0002-5245-4659} \and  Hajo A. Reijers\inst{1}\orcidID{0000-0001-9634-5852}}
\authorrunning{Zuijderwijk et al.}
%
\institute{Utrecht University, Utrecht, The Netherlands\\
\email{a.j.h.zuijderwijk@uu.nl}
\and
Ministry of Infrastructure and Water Management, The Hague, The Netherlands
}

\maketitle              
\begin{abstract}
Government transparency, widely recognized as a cornerstone of open government, depends on robust information management practices. Yet effective assessment of information management remains challenging, as existing methods fail to consider the actual working behavior of civil servants and are resource-intensive. Using a design science research approach, we present the \textit{Transparency Anti-Pattern Assessment System} (TAPAS) — a novel, data-driven methodology designed to evaluate government transparency through the identification of behavioral patterns that impede transparency. We demonstrate TAPAS's real-world applicability at a Dutch ministry, analyzing their electronic document management system data from the past two decades. We identify eight transparency anti-patterns grouped into four categories: \textit{Incomplete Documentation}, \textit{Limited Accessibility}, \textit{Unclear Information}, and \textit{Delayed Documentation}. We show that TAPAS enables continuous monitoring and provides actionable insights without requiring significant resource investments.

 \\


\end{abstract}

\section{Introduction}

Transparency is widely recognized as a cornerstone of open government \cite{meijer2012} and key to digital government success \cite{busch2024theory}. It enables citizens to see and consequently monitor the inner workings of public institutions through access to relevant information. Today, transparency is considered ``a universal benchmark of good governance'' \cite{pippidi_2022}, as reflected in the adoption of Freedom of Information (FOI) laws by 191 countries \cite{freedominfo}. The rise of e-government has enabled transparency by making government information more accessible and shareable than ever before. However, transparency has simultaneously become significantly more complex, with modern government agencies generating unprecedented volumes of data across numerous interconnected systems \cite{cerrillo2021}.

This complexity highlights the need for robust information management. The relationship between transparency and information management practices is fundamental: information management allows for transparent working; without proper documentation, storage and accessibility of information, achieving transparency may be very challenging. Like a library without a catalog system or organized shelves, where finding specific books becomes virtually impossible, governments without robust information management cannot publish relevant information upon request.

In practice, governments evaluate the state of their information management by conducting self-assessment surveys, measuring their progress along progressive stages of capability using a maturity model \cite{proenca2016,thomas2019}. For example, the Dutch government conducts annual surveys for all organizational units (e.g., ministries, departments, subdepartments) with more than 500 employees \cite{AnalyseJaarrapportages2023}. Other examples include Australia \cite{StateRecordsSA2025_IMMS} and the UK \cite{UKGovernment2025_DMA}. While providing valuable frameworks for benchmarking capabilities across organizations and raising organizational awareness through participatory evaluation, these self-assessments have several practical and methodological disadvantages. First, they focus on aspects of governance such as policies and organizational structures, or indirect factors such as the culture of openness rather than the actual behavior related to information management. Second, they capture a yearly, static snapshot rather than ongoing practices. This overlooks how events, changing workloads, and organizational interventions influence information management. Third, conducting these surveys, especially on a national level, is time and resource-intensive. Finally, the empirical objectiveness of such assessments is limited; self-assessments inherently involve subjective judgments \cite{mabe1982} while maturity models may inadvertently create an environment where organizations feel encouraged to demonstrate progression toward higher stages \cite{andersen2006}.

To address the disadvantages of self-assessments, we introduce the \textit{Transparency Anti-Pattern Assessment System} (TAPAS) --— a methodology for assessing government transparency by identifying and measuring behavioral patterns in information management. The fundamental innovation of TAPAS lies in its inverse pattern-based approach: rather than attempting to directly measure transparency, which is difficult to quantify objectively, we identify recurring behavior that undermines transparency, i.e., anti-patterns. TAPAS consists of four phases: (1) anti-pattern discovery, (2) data collection, (3) implementation, and (4) monitoring. Within the study context of the Dutch Ministry of Infrastructure and Water Management (IenW), we demonstrate TAPAS's feasibility, and evaluate its practical utility through member checking with information experts. Unlike traditional methods, TAPAS focuses on working behavior, enables continuous monitoring for actionable insights, is cost and time-effective, and provides objective measurements based on system data.

This paper makes three key contributions: (1) a novel, data-driven approach to measuring government transparency through behavioral patterns in information management, i.e., TAPAS, (2) design
knowledge about transparency assessment, including an anti-pattern catalog, i.e., a systematic categorization of transparency-impeding behavior in information management and (3) a demonstration of the feasibility of TAPAS through a large-scale implementation at a Dutch ministry.

The paper proceeds as follows. Sections \ref{sec:background} and \ref{sec:design} cover related work and our design science research method, respectively. Section \ref{sec:design-development} introduces TAPAS, while Section \ref{sec:implementation} demonstrates its application. Section \ref{sec:evaluation} evaluates our results via experts and practitioners. Section \ref{sec:discussion} examines implications and presents transparency measurement design principles. Section \ref{sec:conclusion} summarizes contributions, implementation guidelines, and future research directions.

\section{Background}
\label{sec:background}
\subsection{Transparency}
\label{sec:openness-and-transparency}

Transparency is a broad, multi-faceted concept with a long history originating in political science \cite{heald2006}. It is associated with the disclosure of information and the perceived quality of the information \cite{albu2019,cruzromero2023}. Information disclosure by governments occurs through two primary channels: passive and active release \cite{meijer2013}. Passive release operates through FOI legislation that allows individuals to request information, without needing to provide a justification for their request. Active release involves governments proactively publishing information, e.g., in the case of open government data. Transparency can be divided into three components: an observer, observable content, and a method of observation \cite{oliver2004}. Being transparent means making internal procedures visible to outsiders, allowing them to verify that an organization is functioning properly \cite{moser2001}. Based on these definitions, Meijer \cite{meijer2013} defines transparency as ``the availability of information about an actor that allows other actors to monitor the workings or performance of the first actor.'' Citizens are the (primary) observers, the government is the observee, and the methods of observation are active or passive information release. Based on this notion, we define civil servants \textit{working transparently} as facilitating the availability of complete, accurate, and timely information. By extension, \textit{working non-transparently} would involve impeding this availability. We note that transparency, in terms of completeness, accuracy and timeliness, exists on a continuum rather than as a binary state.

\subsection{Information Management as a Prerequisite for Transparency}

With digitization and the rise of e-government, the amount of data generated and managed by the government has grown considerably \cite{crivellari2024unveiling}. At the operational level of information management, civil servants handle government information via Electronic Document Management Systems (EDMSs). EDMSs, computer systems that support managing digital information throughout its lifecycle (from creation to destruction), have been widely adopted by government agencies \cite{gani2024}. Information management can be defined as the organized collection, storage, and use of information, including strategies, systems, and practices that help employees handle information \cite{rowley1998}. It also involves maintaining quality of information, ensuring that government information is accurate, trustworthy, and complete \cite{koltay2016data}, as well as proper archiving, preventing information overload.

Robust information management has been explicitly recognized as a prerequisite for achieving transparency in open government policies (e.g., \cite{Drahmann2021,CIOC2016}), although remains largely overlooked in transparency literature. While a few researchers have mentioned this connection (e.g., \cite{cerrillo2021}), to the best of our knowledge, studies do not consider information management separately as a prerequisite of transparency in their analyses. In some conceptualizations of transparency \cite{williams2015}, `information infrastructure' is considered separately from information quantity and quality, but that refers to the infrastructure of providing the outside world with government information, such as data portals or telecommunication channels.

\subsection{Measuring Transparency}
\label{sec:transparency-measurements}

As transparency has a multi-faceted nature, scientific methods for its measurement usually capture specific aspects involving proxies, including the reporting of economic data  \cite{hollyer2014measuring,islam2006transparency}, press freedom \cite{becker2007}, FOI law implementation and performance  \cite{HAZELL2010352}, budget transparency \cite{Seifert01022013}, technical measures of open data quality \cite{zhengEvaluatingGlobalOpen2020}, perceived corruption \cite{TI2024}, and composite indices covering various aspects \cite{kaufmann2005,pippidi_2022,williams2015}. These traditional indices are typically created for comparing governments across different nations as well as for tracking changes over time. In these methods, information management is not taken into account as a factor. Instead, they rely on external indicators or proxies that capture only the observable outcomes of (a lack of) transparency, rather than the underlying processes.

\section{Research Design}
\label{sec:design}
We employ a design science research (DSR) approach \cite{hevner2004,pfeffers2007} to develop, implement, and evaluate TAPAS.  DSR is a research paradigm that aims to generate knowledge by building, iteratively improving and evaluating artifacts \cite{hevner2004}. Artifacts are objects, such as an algorithm, a model or, as in our case, a methodology, that are designed to solve a certain problem. A core idea of DSR is that a research contribution is embedded within the design of the artifact \cite{hevner2004}. We follow the DSR methodology by Pfeffers et al. \cite{pfeffers2007}, which consists of six stages. In the following, we outline these stages and explain how we applied them in our research.

\textbf{1. Problem identification and motivation} requires defining the research problem and justifying the value of a solution. We identified that current assessment methods do not take working behavior into account but instead focus on governance aspects, are time and resource-intensive, capture yearly snapshots, and are prone to subjectivity. Overcoming these disadvantages would allow governments to realize transparency improvements more effectively.

\textbf{2. Definition of the solution objectives} involves defining clear objectives for a solution based on the problem definition. A solution should overcome the disadvantages mentioned in the previous stage. We identify four objectives.
First, the solution should be based on behavior, a more direct measurement (O1). Second, the solution should be significantly less time and resource intensive than current approaches (O2). Third, the solution should enable continuous monitoring for actionable insights (O3) rather than yearly snapshots. Fourth, the solution should enable governments to measure transparency of the organization and smaller organizational units (O4), preserving this capability from existing methods.

\textbf{3. Design and development} is about the creation of the artifact. Our artifact, TAPAS, is a methodology for evaluating government transparency. It takes a pattern-based approach in identifying non-transparent working behavior. It consists of four phases and includes supporting concepts as transparency anti-patterns, detection rules and indicators.

\textbf{4. Demonstration} shows how the artifact solves the problem identified in the first stage. We implemented TAPAS at IenW, applying it to a large EDMS dataset, containing information on all records from 2003 to present. This allowed us to analyze historical patterns of information management behavior and demonstrate the practical applicability of our approach. The implementation showed how the methodology can detect and measure transparency-impeding behavior in practice.

\textbf{5. Evaluation} involves comparing the objectives of a solution to actual observed results from use of the artifact in the previous stage. To that end, we validated our results through member checking \cite{motuslky2021} with information management experts from the ministry.

\textbf{6. Communication} involves sharing the problem, artifact, and results with researchers and practitioners. This paper presents our methodology, implementation, and evaluation results, with concrete recommendations for implementation in government organizations.

\section{Design and Development: TAPAS}
\label{sec:design-development}
Central to our approach is the use of patterns. Patterns help humans make sense of complex systems by identifying recurring solutions to common problems \cite{alexander1977pattern}. This approach has proven valuable across multiple fields as it allows successful solutions to be recognized and reused. An anti-pattern can be seen as a pattern that one should avoid rather than follow. Inspired by the work on workflow patterns \cite{aalst2003workflow}, we apply the idea of anti-patterns in the context of information management --- for analyzing how \textit{well} information management is executed in terms of transparency. We focus on the degree of things that go \textit{wrong}, as identifying failures provides a clearer measuring system than trying to directly measure success.

An \textbf{anti-pattern} represents the underlying behavior that leads to reduced transparency in information management practices. Each anti-pattern is associated with metrics, i.e., one or more detection rules and indicators. Each anti-pattern must be independently verifiable, allowing multiple anti-patterns to apply to the same document without requiring the detection of other patterns for verification. A \textbf{detection rule} defines the specific conditions under which an anti-pattern can be identified. It has a binary nature, where each anti-pattern instance is clearly countable. Furthermore, detection of anti-patterns must be based on data that has been recorded by some system, with detection being reproducible and not relying on assumptions about missing data. \textbf{Indicators} are continuous measurements that can signal the presence of an anti-pattern. Unlike the binary nature of pattern detection, indicators can assume various values along a defined scale. Indicators can be used as performance indicators on their own. In summary, anti-patterns describe behavior as they occur in practice, detection rules represent their measurable manifestations through binary classification, and indicators provide continuous measurements indicating the existence of the anti-patterns. While we could have referred to detection rules as ``indicative rules'' (as in \cite{suriadi2017}) since the rules are \textit{indicative} of the behavior, we aim to have a clear distinction between detection and indication.

\begin{figure}[t!]
\centering

\resizebox{.5\textwidth}{!}{
\begin{tikzpicture}[
    node distance=1cm,
    box/.style={
        rectangle,
        draw,
        text width=3cm,
        minimum height=1cm,
        align=center,
        fill=white
    },
    result/.style={
        rectangle,
        draw=black,
        dashed,
        text width=3cm,
        minimum height=0.8cm,
        align=center,
        fill=white
    },
     arrow/.style={
        ->,
        line width=1.2pt,
        >=latex, 
        shorten >=2pt,
        shorten <=2pt 
    }
]

\fill[fill=gray!10] (-0.5,1) rectangle (7.5,-6);

\node[text width=3cm, align=center] at (1.5,0.5) {\textbf{Phase}};
\node[text width=3cm, align=center] at (5.5,0.5) {\textbf{Result}};

\node[box] (discovery) at (1.5,-0.5) {1. Anti-pattern discovery};
\node[result] (result1) at (5.5,-0.5) {Anti-patterns, abstract metrics};

\node[box] (data) at (1.5,-2) {2. Data collection};
\node[result] (result2) at (5.5,-2) {Up-to-date and historical data};

\node[box] (implementation) at (1.5,-3.5) {3. Implementation};
\node[result] (result3) at (5.5,-3.5) {Operationalized metrics};

\node[box] (monitoring) at (1.5,-5) {4. Monitoring};
\node[result] (result4) at (5.5,-5) {Actionable insights};

\draw[arrow] (discovery) -- (result1);
\draw[arrow] (discovery) -- (data);
\draw[arrow] (data) -- (result2);
\draw[arrow] (data) -- (implementation);
\draw[arrow] (implementation) -- (result3);
\draw[arrow] (implementation) -- (monitoring);
\draw[arrow] (monitoring) -- (result4);

\draw[dotted, arrow] (data.west) to[bend left=40] (discovery.west);

\draw[line width=0.5pt] (-0.5,1) rectangle (7.5,-6);
\end{tikzpicture}}
\caption{Overview of the TAPAS methodology}
\label{fig:TAPAS-method}
\end{figure}
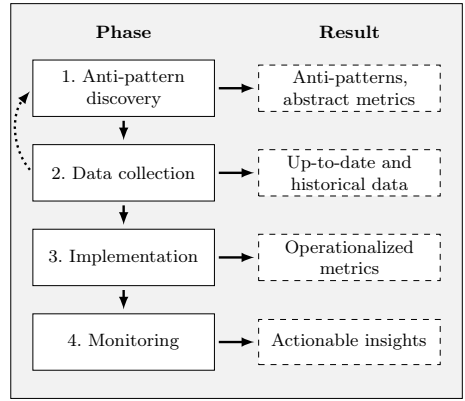

TAPAS consists of four phases: anti-pattern discovery, data collection, implementation, and monitoring. An overview of our methodology is given in Figure \ref{fig:TAPAS-method}. In the (1) \textit{anti-pattern discovery} phase, a list of anti-patterns is obtained through qualitative analysis (i.e., interviews and document review). Interviews should focus on pinpointing specific types of non-transparent working behavior. Internal guidelines on information management (if available) serve as a good starting point. Relevant questions include inquiring whether participants believe they use the EDMS as intended, when they deviate from intended usage, and if they recognize any previously identified anti-patterns. In this phase the ways of detecting or indicating the anti-patterns is also inventoried on a high level. This helps finding relevant data in the next phase.

The (2) \textit{data collection phase} focuses on gathering the data needed for the metrics identified earlier. In e-government, information is typically managed as documents using Electronic Document Management Systems (EDMSs). One key advantage of EDMSs for this purpose is their ability to track data such as where information was stored, when, and by whom, making them a valuable data source. Additionally, other data sources that log information management activities can be used to analyze and reconstruct behavior patterns. We note that the anti-pattern discovery phase and the data collection phase do not need to follow a strict sequential order. The abstract detection rules and indicators can be adjusted based on available data, as there may be multiple viable approaches for detection and indication. Although we present these phases in a specific order, practitioners can approach them iteratively.

In the (3) \textit{implementation phase}, the earlier defined metrics are applied to the collected data. This typically includes data preparation, developing software for calculating the metrics, and organizing the results in a way that is useful for analysis (e.g., by creating a dashboard). In this phase both information specialists and practitioners are consulted to shed light on the data context.

Finally, in the (4) \textit{monitoring phase} we may reap the rewards of the hard work of the previous phases. We can analyze historical trends to gain insight in the development of behavior over time. In this analysis, we can identify long-term patterns (e.g., decreasing frequency of anti-patterns over time) as well as assess the impact of known, significant events, such as migration to a different information system. Through regular measuring at shorter time intervals, such as weekly averages rather than multi-year overviews, we obtain actionable insights at a certain organizational granularity (e.g., the department level). The monitoring phase is not merely about observation but also about active response to the signals provided by the metrics, creating a feedback loop.

\section{Demonstration}
\label{sec:implementation}
We demonstrate our method by an implementation within IenW. We first present the results of the discovery phase in the form of an anti-pattern catalog. Then we discuss data collection, implementation and monitoring for a selection of anti-patterns.

\subsection{Anti-pattern Discovery}
\label{sec:impl-antipatterns}

In this initial phase of TAPAS, we conducted semi-structured interviews with nine participants, all employees of the ministry, in total. The selection of interviewees was driven by two strategies. First, we purposefully sampled participants across different organizational roles to capture diverse perspectives on information management. The sample included process advisors, department heads, FOI request coordinators and information management advisors. Second, we employed snowball sampling, where initial participants suggested additional colleagues whose roles they considered relevant to the study.
Interviews lasted approximately one hour each, were audio-recorded with participant consent, and transcribed and coded afterwards.  We reached theoretical saturation after having spoken to six participants, with no new anti-patterns coming up in subsequent interviews. The questions focused on practices in their usage of the EDMS, the perceived impact of those practices on transparency, and their perceived prevalence.

We identified eight distinct anti-patterns that impede transparency. Table \ref{tab:anti_patterns} provides an overview including the categorization, identifiers, concise descriptions and possibly enlightening icons. Note that we refrain from discussing these anti-patterns in terms of intentionality or legality (as in \cite{pasquier_organization_barriers_2007}). For our purpose, it suffices to point to the lack of transparency, as the intention behind the anti-pattern does not affect its impact. 
The identified anti-patterns can be grouped into four main categories based on how they affect transparency:

\begin{table}[t!]
\scriptsize
\caption{Overview of Transparency Anti-Patterns}

\setlength{\tabcolsep}{8pt} 
\renewcommand{\arraystretch}{1} 

\makebox[\textwidth][c]{
\begin{tabular}{m{0.75cm}m{0.5cm}m{3cm}m{7cm}}
\toprule
\textbf{Icon} & \textbf{Id} & \textbf{Name} & \textbf{Description} \\
\midrule
\multicolumn{4}{l}{\textit{I1. Incomplete Documentation}}\\
\midrule
\includegraphics[height=0.6cm]{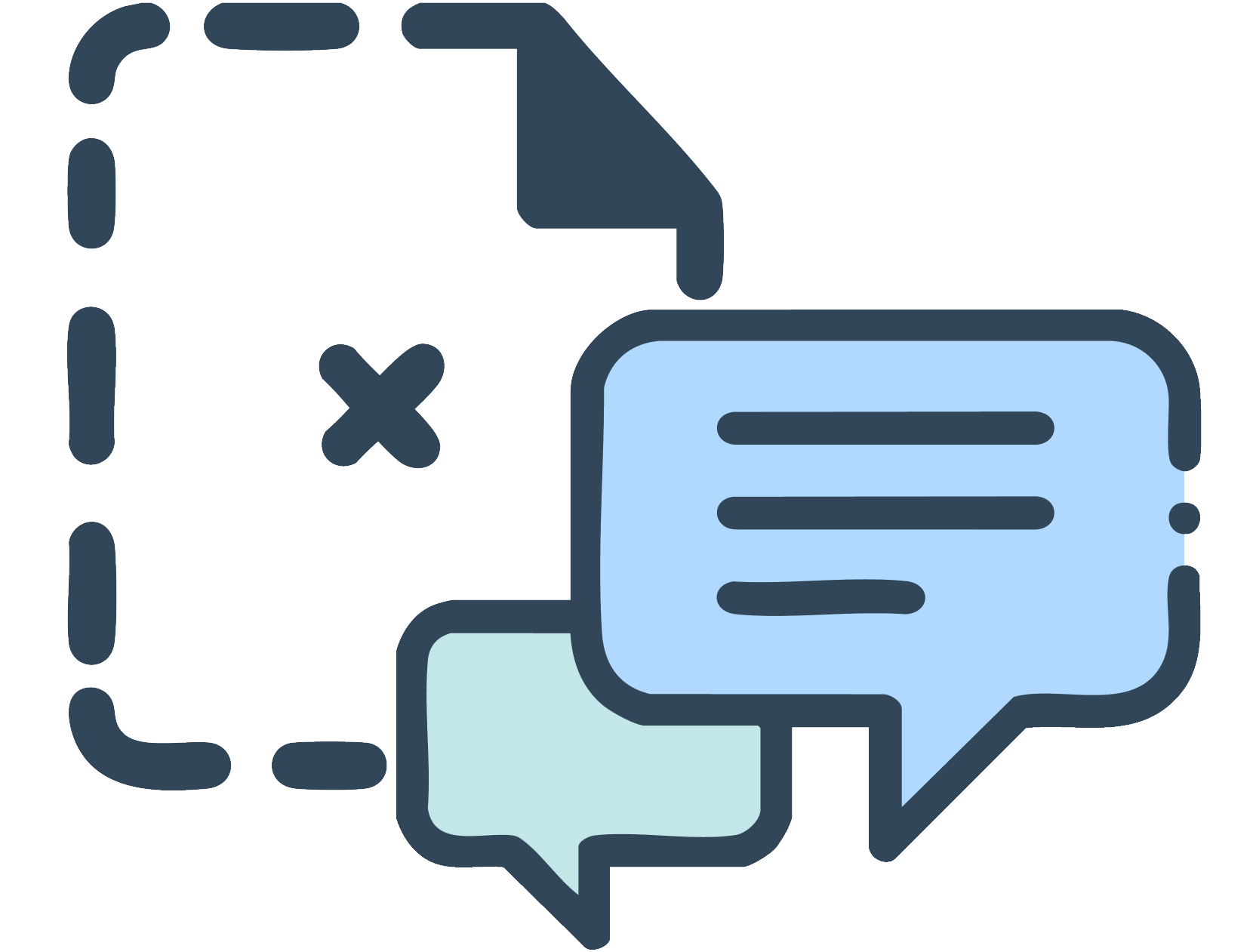} & I1.1  & \makecell[cl]{Documentation \\ Avoidance} & \RaggedRight{Relying on verbal communication or informal channels rather than documenting using designated systems.} \\

\includegraphics[height=0.6cm]{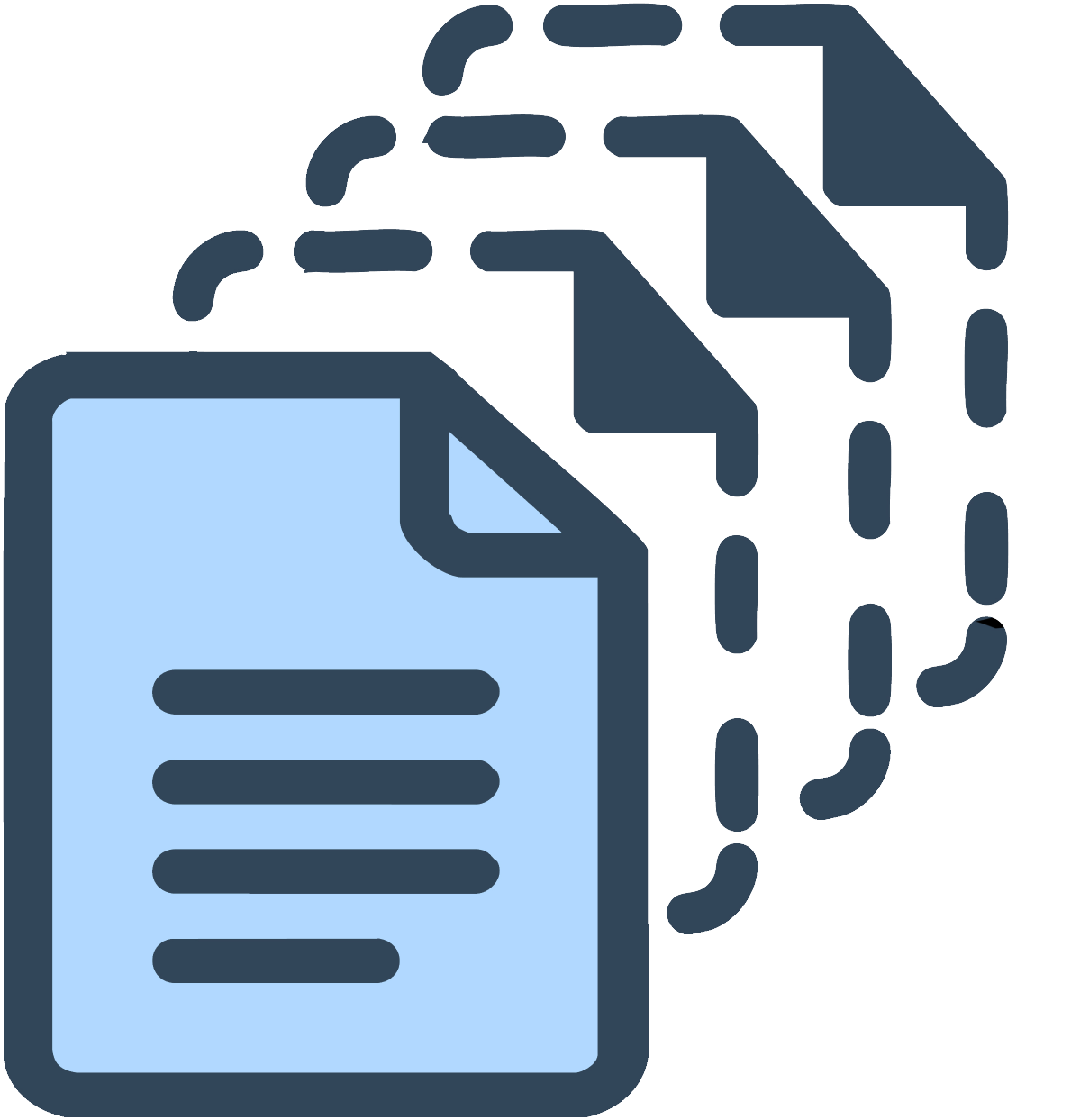} & I1.2 & \RaggedRight{Final Version Only} & \RaggedRight{Only storing the final version of a document without maintaining version history (decision context).} \\
\midrule
\multicolumn{4}{l}{\textit{I2. Limited Accessibility}} \\
\midrule
\includegraphics[height=0.6cm]{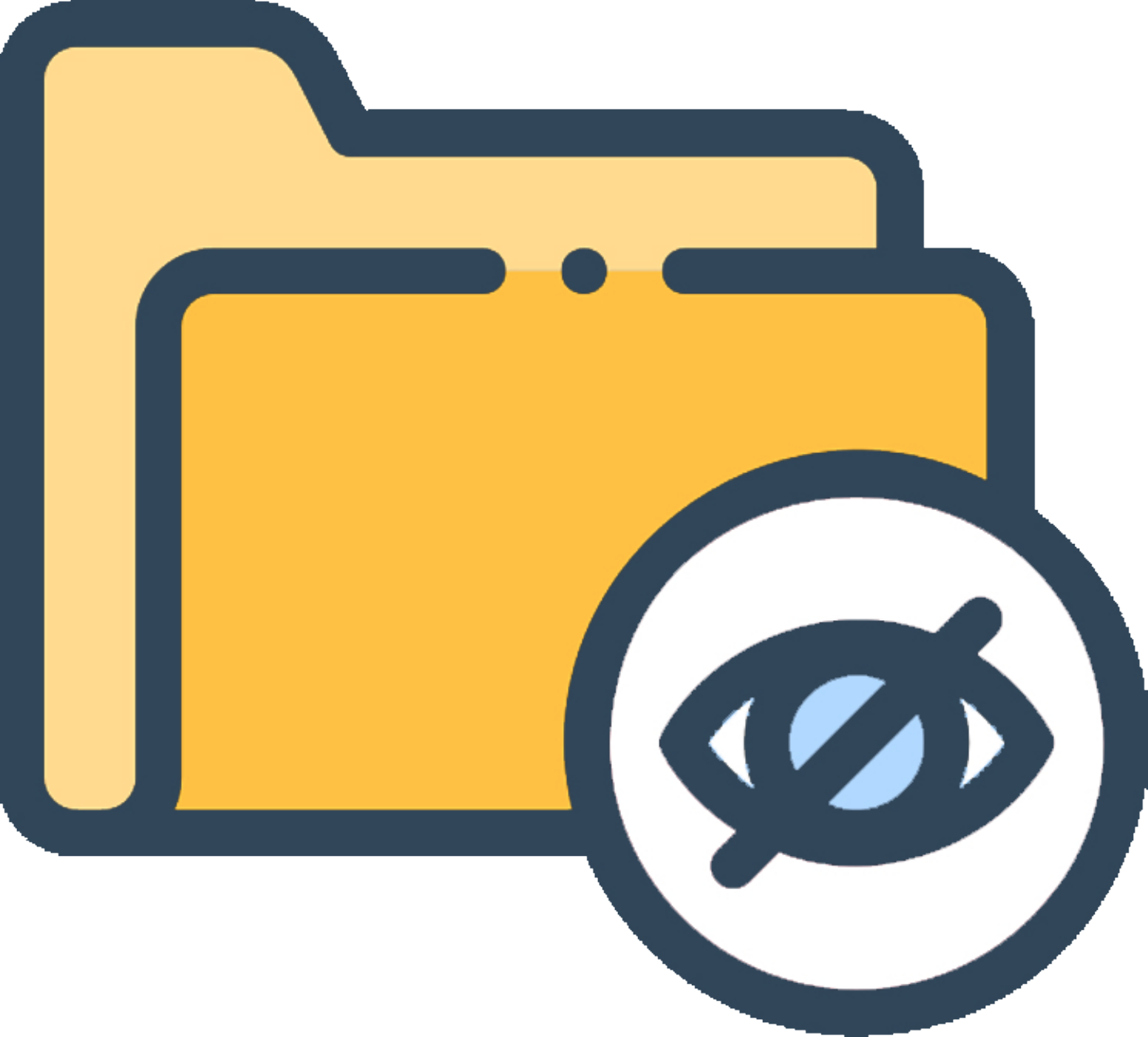} & I2.1 & Inaccessible Storage
 & \RaggedRight{Storing work-related documents in locations not accessible to others within the organization.} \\

\includegraphics[height=0.45cm]{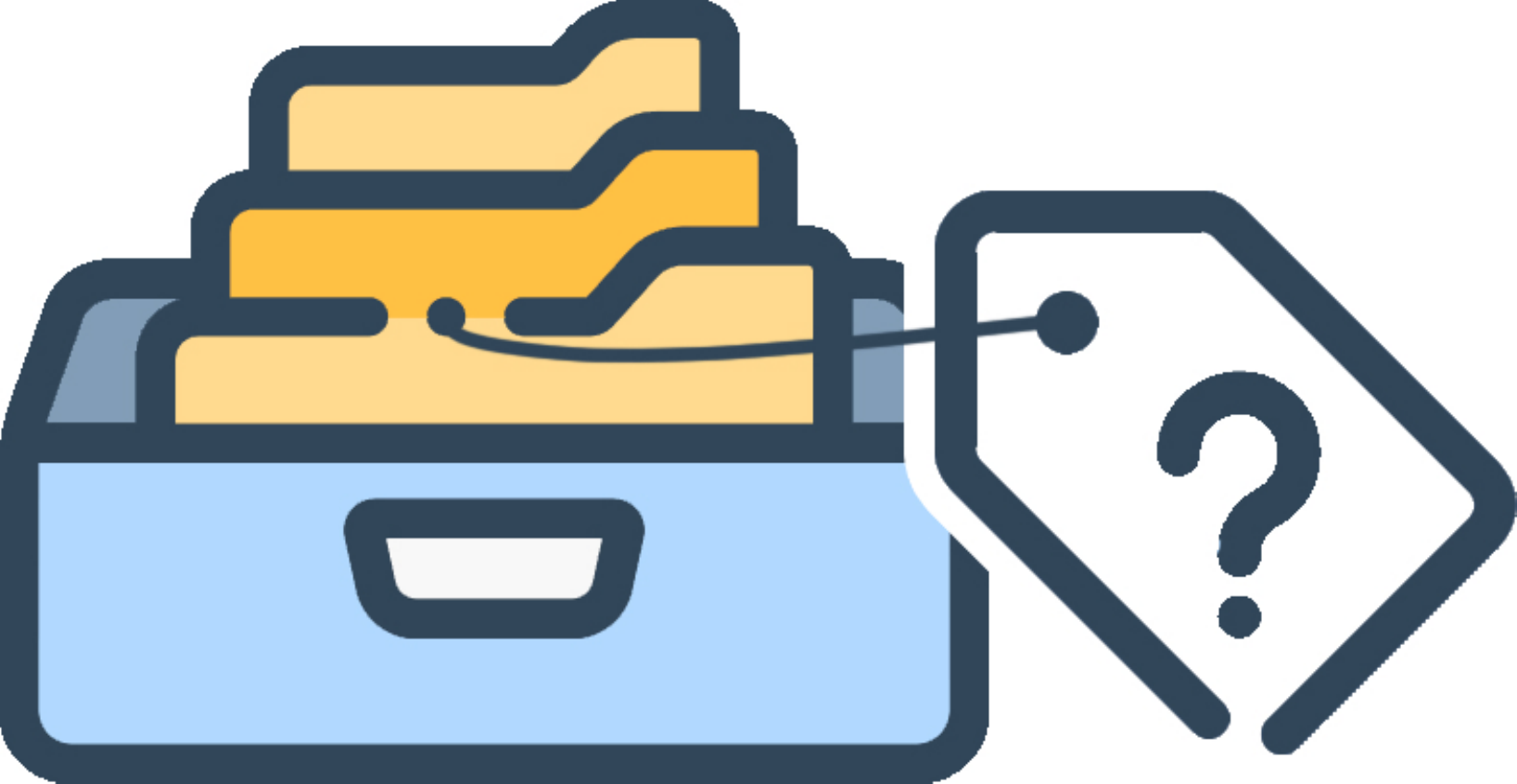}& I2.2 & \makecell[cl]{Non-compliant \\ Structure} & \RaggedRight{Not following the required folder hierarchy, placing documents or folders at incorrect places.} \\

\includegraphics[height=0.5cm]{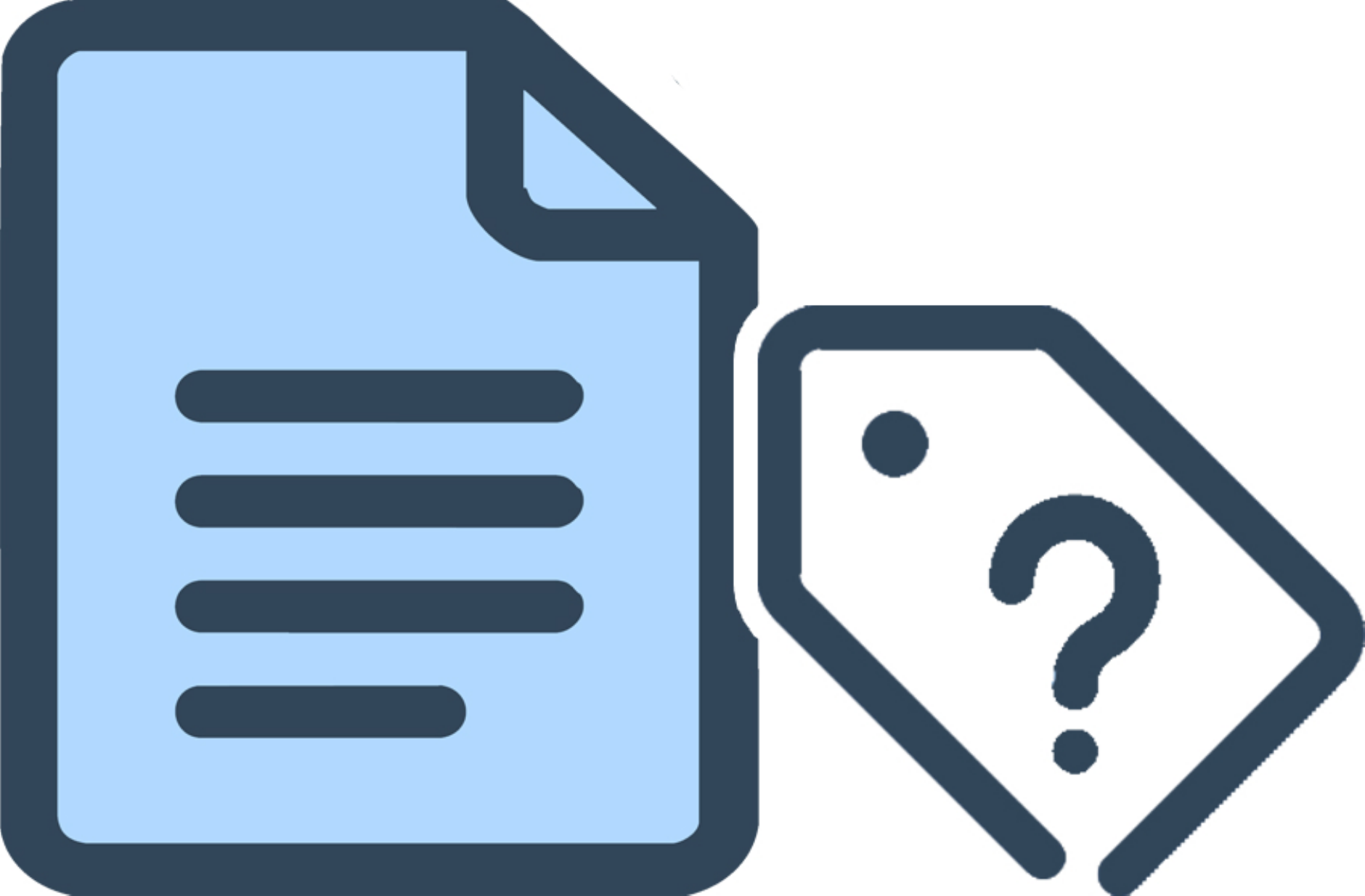} & I2.3 & \RaggedRight{Non-standard Naming} & \RaggedRight{Using unclear, personal, or non-standard naming conventions for files and folders.} \\
\midrule
\multicolumn{4}{l}{\textit{I3. Unclear Information}}\\
\midrule
\includegraphics[height=0.55cm]{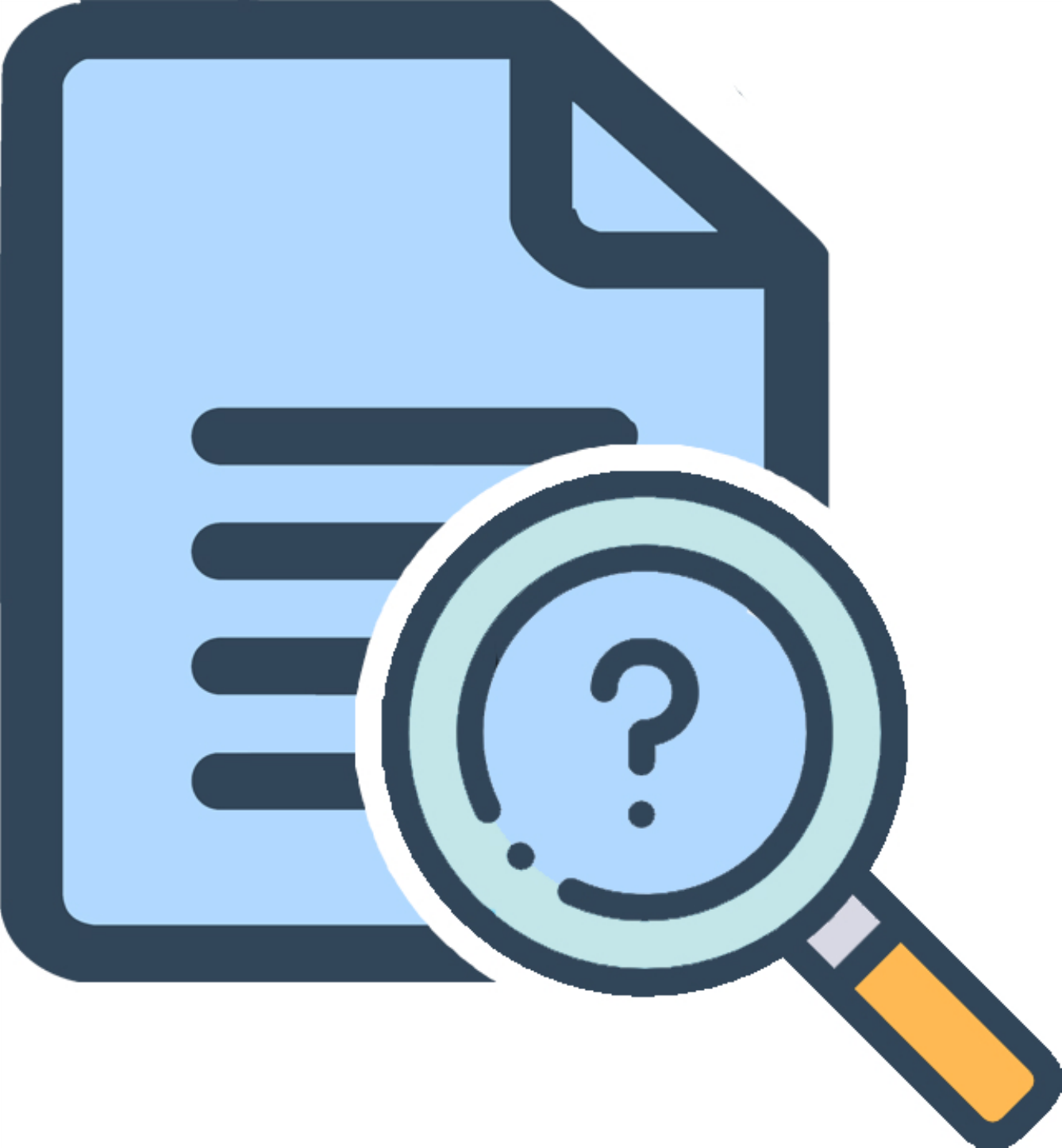}& I3.1 & \raggedright{Opaque Language} & \RaggedRight{Using undefined abbreviations, jargon, or technical terms without explanation.} \\
\midrule
\multicolumn{4}{l}{\textit{I4. Delayed Documentation}}\\
\midrule
\includegraphics[height=0.6cm]{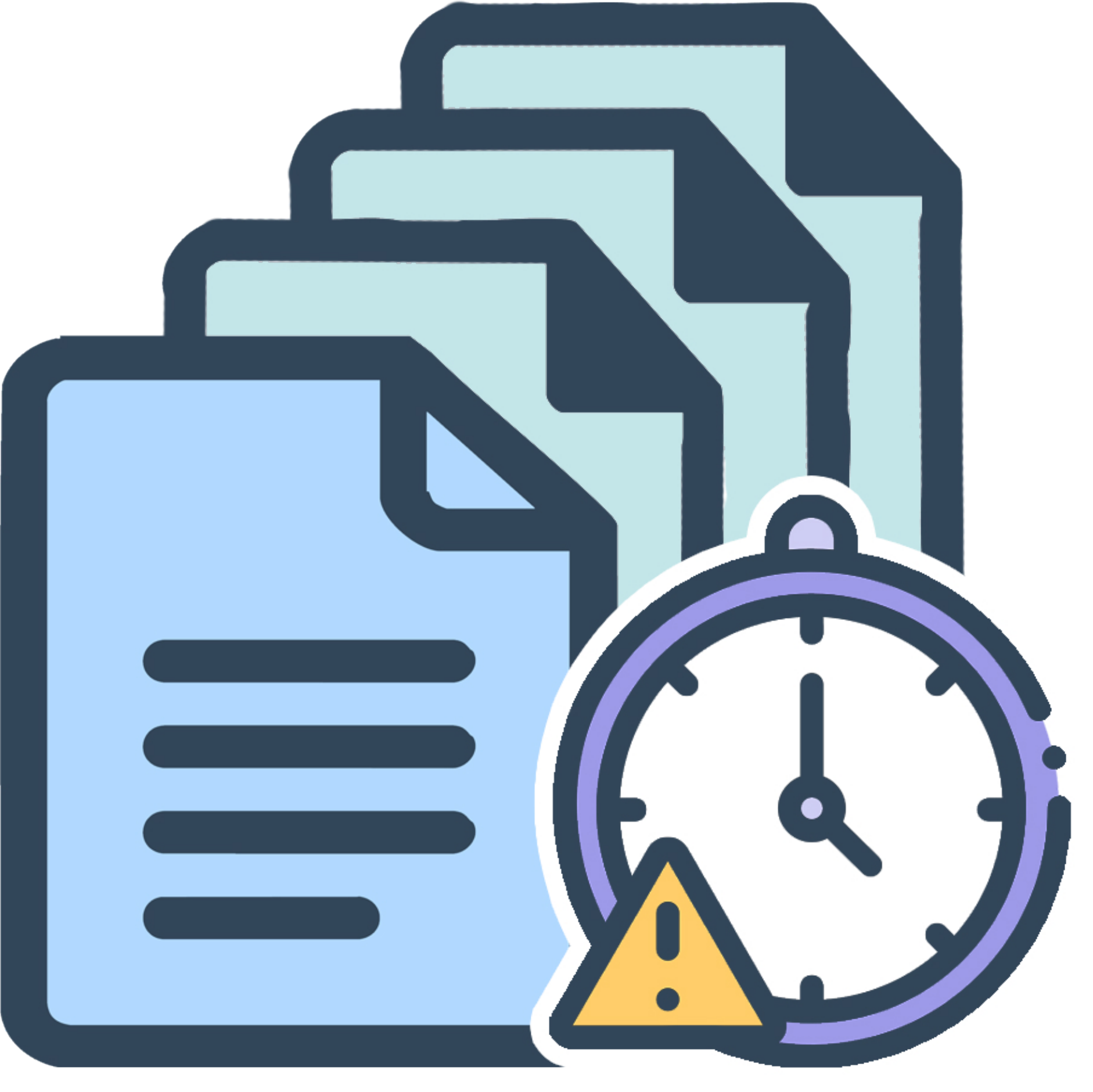} & I4.1 & \raggedright{Batch Documentation} & \RaggedRight{Creating multiple related documents in a single batch after extended periods.} \\

\includegraphics[height=0.55cm]{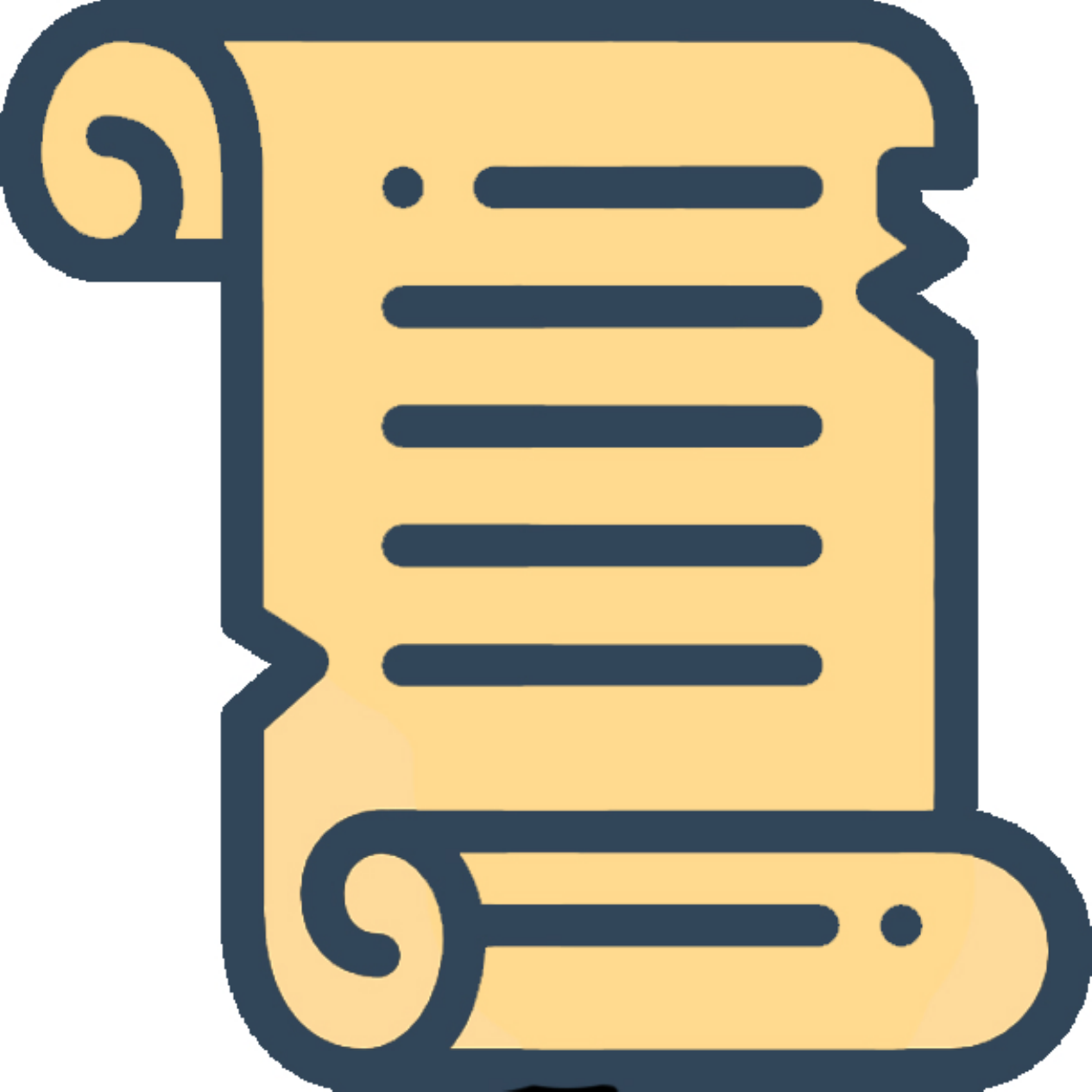} & I4.2 & \makecell[cl]{Abandoned \\ Documentation} & \RaggedRight{Leaving documents or folders unarchived despite long periods of inactivity.} \\
\bottomrule
\end{tabular}}
\label{tab:anti_patterns}
\end{table}

\begin{enumerate}
    \item[I1] \textbf{Incomplete Documentation}: These patterns result in missing information, either because it was never documented or because important context and history were not preserved;

\item[I2] \textbf{Limited Accessibility}: These patterns make it difficult or impossible to find or access information, even when it exists somewhere in the organization;

\item [I3] \textbf{Unclear Information}: These patterns make information difficult to understand even when it can be found, due to unclear or ambiguous language use;

\item[I4] \textbf{Delayed Documentation}: These patterns involve timing issues where information is not documented or properly archived when it should be, creating risks of information loss, periods of inaccessibility, or information overload.
\end{enumerate}
For each identified anti-pattern, we provide a concrete example of an occurrence, a supporting quote from the interviews, the detection rules and indicators in Appendix \ref{appendix:anti-patterns}.

\subsection{Data Collection}
For our explorative demonstration of the use of the anti-patterns, we present results of the implementation of \textit{Final Version Only}, \textit{Inaccessible Storage}, \textit{Batch Documentation} and \textit{Abandoned Documentation}. We made this selection based on the direct availability of the data needed (e.g., we did not obtain file names) and because the goal of our implementation is to demonstrate feasibility. We have gathered the necessary data extracted from the EDMS and preprocessed it to obtain records of documents and folders, the number of files within each folder, document types, creation and last updated timestamps, creator IDs, associated organizational units, revision numbers (total count of document versions), and archival states. The data encompasses a comprehensive dataset containing records of all documents managed across the entire ministry and its subordinate agencies, spanning a period of two decades. We will not delve further into the preprocessing steps, as they vary greatly depending on each organization's unique dataset, practices, and systems. 

\subsection{Implementation}

After we collected the data, we implemented the metrics per anti-pattern. For \textit{Final Version Only}, we implemented the indicator `single version percentage', which measures the proportion of documents with exactly one revision. We have not implemented a detection rule, as this requires understanding which documents are required to have multiple versions  (e.g., based on document type), necessitating a more in-depth analysis. However, in our implementation of the indicator we have filtered out the document types that were bound to have exactly one version (e.g., stored email messages).

For \textit{Inaccessible Storage}, we have implemented the indicator `percentage of empty folders,' which calculates the proportion of folders that contain no child records (files or subfolders) at a specific time $t$. A folder is considered \textit{empty} if it has no children created on or before the time of measurement $t$. The rationale behind this indicator is that empty project folders suggest that files are stored elsewhere. \textit{Detecting} this anti-pattern is challenging because it requires information about these alternative storage locations, which are often inaccessible. Despite this, data on personal disk space usage should be available within the organization.

For \textit{Batch Documentation}, we have implemented a detection rule that identifies a ``batch'' as a large number of documents created by the same person within a short timeframe. Specifically, we define a batch as occurring when a single user uploads 50 or more documents within a 30-minute window, thresholds chosen to identify both rapid automated and deliberate manual uploads of significant size. Each document is counted only once within a batch, and for a new batch to be recognized, we require a gap of at least 30 minutes after the last document from the previous batch. The indicator `documents per person' also takes the size of these batches into account, enhancing our understanding of individual upload patterns.

Finally, for the \textit{Abandoned Documentation} anti-pattern, we implemented a detection rule that identifies a record as \textit{abandoned} if it is neither archived nor modified for an extended period. We set this period to one year based on internal guidelines. If a file is abandoned at time 
$t$, it will no longer be classified as such at time $t+2$ if it is archived at time $t+1$. We normalized the occurrences of this anti-pattern, as this is highly dependent on the total number of documents. This normalization makes it easier to compare across different time periods and organizational units.

\begin{figure}[t]
    \centering
    \includegraphics[width=0.95\textwidth]{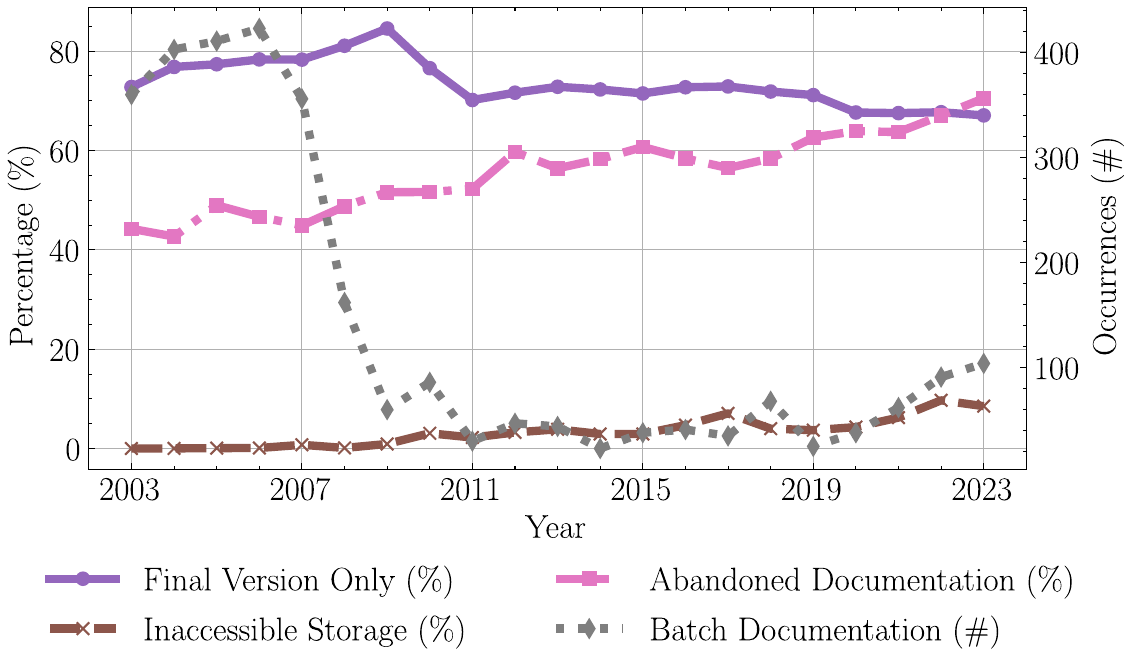}
    \vspace*{-0.25cm}
    \caption{Indicators and detection occurrences averaged over the whole organization}
    \label{fig:averages}
       
\end{figure}

\begin{figure}[t]
    \centering
  
    \makebox[\textwidth][c]{\includegraphics[width=1.25\textwidth]{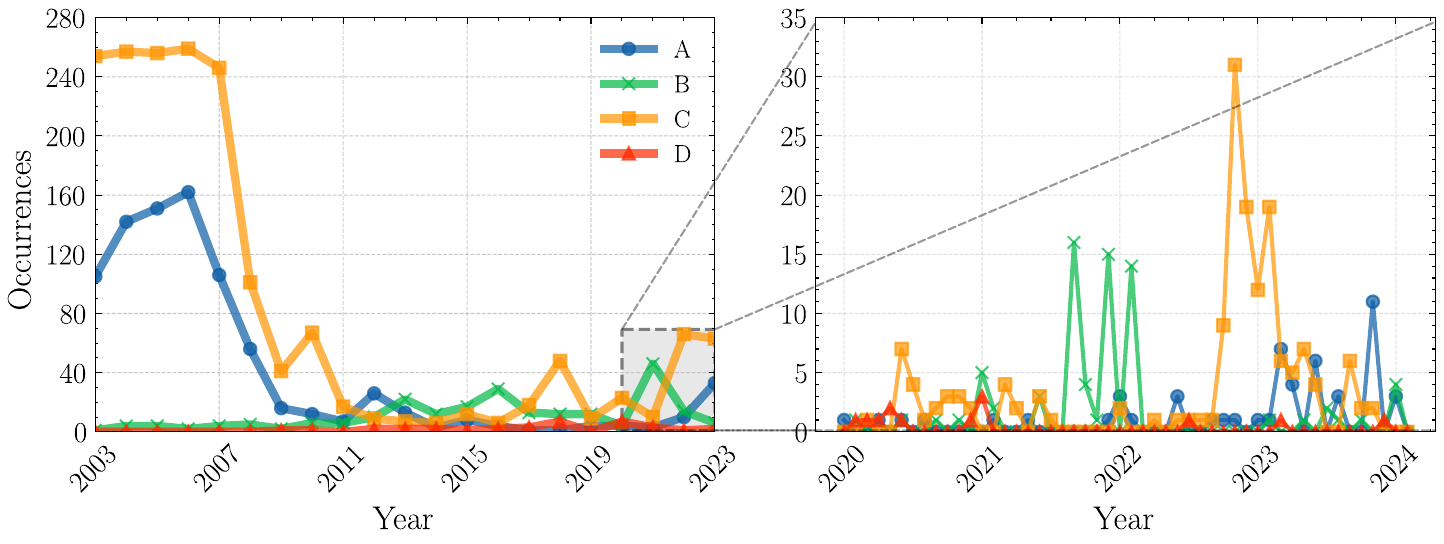}}
    \vspace*{-0.5cm}
    \caption{Detection of \textit{Batch Documentation} yearly (left) and monthly (right) per organizational unit}
    \label{fig:i4.1_detection}
\end{figure}

Figure \ref{fig:averages} presents the yearly averaged metrics across the entire organization over the full timespan of the dataset. To illustrate a more detailed analysis, we selected four representative organizational units within the ministry and plotted the occurrences of \textit{Batch Documentation}  in Figure \ref{fig:i4.1_detection}. These organizational units were pseudonymized to maintain confidentiality. The left plot displays yearly averages, while the right plot shows monthly averages for the last three years. This figure highlights both long-term trends and recent fluctuations.


\subsection{Monitoring}
The scope of this paper is limited to demonstrating the feasibility of TAPAS. Therefore, we have not comprehensively monitored these values over an extended period or actively intervened with the goal of reducing the occurrences of anti-patterns. However, we can identify trends by examining the historical data. Notably, Figure \ref{fig:averages} displays a sharp decrease in the number of batch documentations. This decline is particularly evident in organizational units A and C (see Figure \ref{fig:i4.1_detection} left). Given that the ministry became fully digitalized leading up to 2012, we suspect that these batches may originate from migrations from other  systems. The other patterns show a steady trend upwards (\textit{Abandoned Documenation}, \textit{Inaccessible Storage}) or downwards (\textit{Final Version Only}).

In the more detailed, monthly summed view on the right of Figure \ref{fig:i4.1_detection} we observe three distinct peaks for unit B that drop to zero between occurrences. This might suggest that the previous month's work was entered into the system as a single batch. Additionally, we can observe peaks for unit C that persist for longer durations, indicating sustained activity periods rather than singular events. These observations could serve as topics for discussions with unit C.

\section{Evaluation}
\label{sec:evaluation}

We conducted member checking \cite{motuslky2021} with six information management experts to assess the practical utility and effectiveness of TAPAS. Initially, we presented our results in a group setting, soliciting feedback on the methodology, anti-patterns, and indicators. Following this, we conducted a one-on-one interview, lasting approximately 90 minutes, to delve deeper into the details. The experts recognized the anti-patterns as relevant and reflective of real-world scenarios. They appreciated the methodology's focus on working behavior, which provided a clearer picture of how information management practices impact transparency. The indicators and detection rules were deemed effective, with suggestions for additional indicators. Experts particularly valued the detailed, zoomed-in insights for monitoring. TAPAS was favored over existing information management assessments due to its direct feedback and oversight capabilities. Further context on the occurrence anti-patterns was also provided. For example, for some divisions it is known that they batch document their information in particular cases for practical reasons.  

Our implementation at the Dutch ministry demonstrates that TAPAS achieves the objectives it was designed to achieve. First, TAPAS identifies non-transparent working behavior (O1) through the identification and analysis of anti-patterns, which were collected through interviews within the organization. The implementation successfully operated efficiently with minimal resource investments (O2), thanks to automated measuring that, once implemented, require little ongoing effort compared to traditional survey-based self-assessments. While initial setup requires an investment in anti-pattern discovery and implementation, the recurring costs are significantly lower than manual assessment methods. TAPAS established continuous monitoring (O3), providing actionable insights through regular assessments and comparative analyses across departments. Finally, the system effectively measured at different organizational granularity (O4), as the EDMS maintained records of document ownership across all organizational units.

\section{Discussion}
\label{sec:discussion}

\subsection{Theoretical Implications}
Our study offers a new perspective on transparency by focusing on the behavior that underlies information management. Traditional transparency literature defines the concept as an information relation between government and external stakeholders, measured through indirect indicators like FOI compliance \cite{HAZELL2010352}, or proxies of transparency like the perception of corruption \cite{TI2024}. These measures focus on external effects or proxies of transparency rather than internal processes enabling transparency. We reconceptualize transparency not merely as a state of disclosure but as a capability enabled or hindered by working behavior in information management. Our approach differs by focusing on the actual working behavior of civil servants who manage information, particularly in how they store and document government information. This addresses both a theoretical gap in transparency literature, which rarely considers information management separately as a prerequisite condition, and a practical gap in how governments internally assess their transparency capabilities. Our inverse approach provides a more concrete measurement framework that connects day-to-day information practices with transparency outcomes. 
In doing so, we provide insight into the dynamic nature of \textit{invisible work} by government employees, i.e., the behavior that takes place behind the formal work \cite{star1999layers}. Data is fragmented across different systems and government workers adapt their use of the prescribed technology \cite{crivellari2024unveiling}. To the best of our knowledge, our methodology is the first to discover this behavior from data and across time.

\subsection{Practical Implications}

While our demonstration focused on a national government, the methodology's anti-pattern approach makes it adaptable to various government levels, including local governments. The pattern-based framework allows organizations to adopt the core methodology while tailoring specific detection rules to their unique information systems and organizational contexts. The framework also offers a practical advantage in its technology-agnostic design --- organizations can implement TAPAS using existing analytics capabilities without requiring specialized transparency assessment tools or expensive consultancy services.

For applying TAPAS in practice, we recommend forming working groups with expertise in information management (policy), data analysis, and privacy. Organizations should establish protocols for responding to identified anti-patterns and create feedback loops between monitoring results and organizational improvements. Particular attention should be paid to engaging information management specialists and end users in the interpretation of results, as contextual understanding is crucial for distinguishing between benign variations and genuine transparency impediments.

\subsection{Limitations}

Measuring a lack of transparency is inherently challenging due to information gaps. While our pattern-based approach provides an objective framework, rule-based detection is not infallible and can result in false positives and negatives. For instance, a false positive might occur when the system flags a legitimate bulk upload of consultation responses as \textit{Batch Documentation}, when in fact the civil servant is efficiently processing related documents rather than avoiding timely documentation. Conversely, a false negative could occur with \textit{Documentation Avoidance} when a discussion is neither planned in the agenda nor documented, presenting the appearance of compliance while still impeding transparency. In general, false positives are preferable because it enables governments the identify them as such. Furthermore, the current implementation is limited in scope, focusing on demonstrating feasibility by implementing a selection of metrics. Longitudinal studies applying TAPAS across multiple years with active interventions would provide further insights into its effectiveness.

Beyond the practical measurement challenges, a limitation of this study is its narrow conceptualization of transparency that is based on information disclosure. Future research could benefit from incorporating a \textit{performativity} lens \cite{albu2019}, recognizing transparency as a complex social process where communication practices, power dynamics, and technologies do not just reveal existing organizational behaviors but actively shape and transform them. 

\section{Conclusion}
\label{sec:conclusion}
This paper introduced TAPAS, a novel methodology for assessing government transparency through behavioral pattern detection in information management. Our research makes three primary contributions. First, we developed a data-driven approach shifting from measuring external outcomes to internal practices. TAPAS enables continuous monitoring of organizational transparency capabilities. This addresses a critical gap in both literature and practice, where existing methods primarily rely on external indicators or resource-intensive self-assessments, respectively.
Second, we established a catalog of transparency anti-patterns. The eight identified anti-patterns, grouped into four impact categories, provide a structured framework for understanding and measuring impediments to transparency and serves as a useful starting point for practitioners. Third, we demonstrated the practical applicability of our methodology through large-scale implementation at IenW.

Future research could deepen this work by focusing on individual anti-patterns, developing more sophisticated analysis methods for specific issues such as reconstructing policy development timelines by comparing document versions, clarifying bureaucratic language using internal data sources, or generating compliant file names. Additionally, future research could explore the relationship between anti-pattern metrics and existing transparency indices, or operational measures such as FOI request handling performance.

\appendix
\section{Anti-Patterns Catalog}
\label{appendix:anti-patterns}
\begin{credits}
\textbf{I1.1 Documentation Avoidance.} \textit{Detection Rule:} Calendar event exists requiring documentation, but documentation is absent. \textit{Indicators:} Documentation per calendar event; documentation per project. \textit{Example:} Two civil servants making decisions at the coffee corner without recording notes. \textit{Quote:} ``Suppose we need to decide something together and we're sitting here together. But we don't create a meeting report.''

\textbf{I1.2 Final Version Only.} \textit{Detection Rule:} Document exists requiring iterative development, but no previous versions found. \textit{Indicators:} Mean version count; Single version percentage. \textit{Example:} Only final approved policy version uploaded after extensive stakeholder discussions. \textit{Quote:} ``Sometimes you really only see the last document, because that's what they put in [EDMS]. And we actually don't know at all what preceded it.''

\textbf{I2.1 Inaccessible Storage.} \textit{Detection Rule:} Work document exists outside EDMS with no EDMS copy despite requirements. \textit{Indicators:} Alternative storage usage; percentage of empty folders. \textit{Example:} Project team storing documents on personal network drives, unavailable when requested years later. \textit{Quote:} ``There are certain teams that have to collaborate so much with people from outside or other ministries that they prefer to use a collaboration space rather than [EDMS].''

\textbf{I2.2 Non-compliant Structure.} \textit{Detection Rule:} Record exists in location violating structural rules. \textit{Indicators:} Number of documents per lowest level folder, distribution of the files over folders. \textit{Example:} Project files stored in temporary folders instead of designated project folders. \textit{Quote:} ``One folder with a thousand documents, a `catch-all'.''

\textbf{I2.3 Non-standard Naming.} \textit{Detection Rule:} Record name lacks required descriptive elements. \textit{Indicators:} Record name length. \textit{Example:} Folder called ``Barry'' for Afsluitdijk maintenance 2024. \textit{Quote:} ``Sometimes people only put in a year, for example. (...) like yeah a year is nice, `2013', but what is it actually about?''

\textbf{I3.1 Opaque Language.} \textit{Detection Rule:} Document contains undefined abbreviations not in approved list. \textit{Indicators:} Jargon density (special terms/total words). \textit{Example:} A report using numerous internal acronyms without explanation. \textit{Quote:} ``Certain professional jargon becomes outdated quickly. So two years ago, everyone knew exactly what that one abbreviation meant... Someone who takes over the next case has no idea.''

\textbf{I4.1 Batch Documentation.} \textit{Detection Rule:} Large number of documents created within relatively short timeframe by the same person. \textit{Indicators:} Created documents per person per day. \textit{Example:} Retiring civil servant uploading years of documents just before leaving. \textit{Quote:} ``If your project lasts 1.5 years... Then the project is completed. Then you think `oh yeah, I still need to archive some emails.' ''

\textbf{I4.2 Abandoned Documentation.} \textit{Detection Rule:} Unarchived document or folder with no modifications for over a long period. \textit{Indicators:} the total number of archived documents, the total number of to be destroyed documents \textit{Example:} Keeping outdated project documentation in active folders. \textit{Quote:} ``Yes, then it will never end up in the archive. No, and then no retention or destruction period will ever be assigned to it. And then it will stay here for eternity.''
\end{credits}

 \bibliographystyle{splncs04}
\bibliography{main}

\end{document}